\RequirePackage{fix-cm}
\documentclass[smallextended]{svjour3}       
\smartqed  
\usepackage{graphicx}
\graphicspath{{figures/}{images/}}
\usepackage{multirow}
\usepackage{amsmath}
\usepackage{amssymb}
\usepackage{bm}
\usepackage[misc]{ifsym}
\usepackage{algorithm}
\usepackage{algorithmic}
\usepackage[caption=false,font=footnotesize]{subfig}
\usepackage[bookmarks=true, bookmarksnumbered=true, bookmarksopen=true,
  unicode=true, colorlinks=true,
  linkcolor=blue,citecolor=blue,filecolor=blue,urlcolor=blue,
  pdfstartview=FitH
]{hyperref}
\usepackage{url}
\usepackage{tikz}
\usetikzlibrary{calc}
\usepackage[nodisplayskipstretch]{setspace} 
\newcommand{\tikzmark}[1]{\tikz[overlay,remember picture] \node (#1) {};}
\newcommand{\DrawBox}[4][]{%
  \tikz[overlay,remember picture]{%
    \coordinate (TopLeft)     at ($(#2)+(-0.2em,0.9em)$);
    \coordinate (BottomRight) at ($(#3)+(0.2em,-0.3em)$);
    \path (TopLeft); \pgfgetlastxy{\XCoord}{\IgnoreCoord};
    \path (BottomRight); \pgfgetlastxy{\IgnoreCoord}{\YCoord};
    \coordinate (LabelPoint) at ($(\XCoord,\YCoord)!0.5!(BottomRight)$);
    \draw [red,#1] (TopLeft) rectangle (BottomRight);
    \node [below, #1, fill=none, fill opacity=1] at (LabelPoint) {#4};
  }
}
\usepackage
	{todonotes}

\DeclareMathOperator*{\argmax}{arg\,max}
\journalname{Multimedia Tools and Applications}
\begin{document}

\title{Remembering winter was coming
}
\subtitle{Character-oriented video summaries of TV series}


\author{Xavier~Bost \and Serigne~Gueye \and Vincent~Labatut \and
  Martha~Larson \and Georges~Linar\`es \and Damien~Malinas \and
  Rapha\"el~Roth}


\institute{Xavier Bost (\Letter) \at
  \textit{Orkis}, 13290 Aix-en-Provence, France \\
  Tel.: 0033-4-42612320 \\
  \email{xbost@orkis.com}
  \and
  Xavier~Bost \and Serigne~Gueye \and Vincent~Labatut \and Georges~Linar\`es \at
  \textit{Laboratoire Informatique d'Avignon}, Avignon University, 84000 Avignon, France \\
  \email{\{firstname\}.\{lastname\}@univ-avignon.fr}
  \and
   Damien~Malinas \and Rapha\"el~Roth \at
   \textit{Centre Norbert Elias}, Avignon University, 84000 Avignon \\
   \email{\{firstname\}.\{lastname\}@univ-avignon.fr}
   \and
   Martha~Larson \at
   \textit{Intelligent Systems Department}, Delft University of Technology, The Netherlands \\
   \textit{Centre for Language Studies}, Radboud University Nijmegen, The Netherlands \\
   \textit{Institute for Computing and Information Sciences}, Radboud University Nijmegen, The Netherlands \\
   \email{m.a.larson@tudelft.nl}
}

\date{Received: date / Accepted: date}

\maketitle

\begin{abstract}
  Today's popular \textsc{tv} series tend to develop continuous,
  complex plots spanning several seasons, but are often viewed in
  controlled and discontinuous conditions. Consequently, most viewers
  need to be re-immersed in the story before watching a new
  season. Although discussions with friends and family can help, we
  observe that most viewers make extensive use of summaries to
  re-engage with the plot. Automatic generation of video summaries of
  \textsc{tv} series' complex stories requires, first, modeling the
  dynamics of the plot and, second, extracting relevant sequences. In
  this paper, we tackle plot modeling by considering the social
  network of interactions between the characters involved in the
  narrative: substantial, durable changes in a major character's
  social environment suggest a new development relevant for the
  summary. Once identified, these major stages in each character's
  storyline can be used as a basis for completing the summary with
  related sequences. Our algorithm combines such social network
  analysis with filmmaking grammar to automatically generate
  character-oriented video summaries of \textsc{tv} series from
  partially annotated data. We carry out evaluation with a user study
  in a real-world scenario: a large sample of viewers were asked to
  rank video summaries centered on five characters of the popular
  \textsc{tv} series \textit{Game of Thrones}, a few weeks before the
  new, sixth season was released. Our results reveal the ability of
  character-oriented summaries to re-engage viewers in television
  series and confirm the contributions of modeling the plot content
  and exploiting stylistic patterns to identify salient sequences.
\end{abstract}

\vspace{2mm}

\noindent \textcolor{red}{\textbf{Cite as:}\\X.~Bost, S.~Gueye, V.~Labatut, M.~Larson, G.~Linar\`es, D.~Malinas, R.~Roth\\\href{https://doi.org/10.1007/s11042-019-07969-4}{Remembering winter was coming: Character-oriented video summaries of TV series.}\\Multimedia Tools and Applications, 2019, 78 (24), pp. 35373-35399.\\doi: \href{https://doi.org/10.1007/s11042-019-07969-4}{10.1007/s11042-019-07969-4}}

\section{Introduction}
\label{sec:intro}

These past ten years, \textsc{tv} series became increasingly popular:
for more than half of the people we polled in our user study
(described in Subsection~\ref{subsec:user_study}), watching
\textsc{tv} series is a daily occupation, and more than 80\% watch
\textsc{tv} series at least once a week. Such a success is probably in
part closely related to modern media. The extension of
high-speed internet connections led to unprecedented viewing
opportunities: streaming or downloading services give control to the
user, not only over the contents he will watch, but also over the
viewing frequency.

The typical dozen of episodes that a \textsc{tv} series season
contains is usually watched over a much shorter period of time than
the usual two months it is aired on television: for 41\% of the
people we polled, a whole season (about 10 hours of viewing in average) is
watched in only one week, with 2-3 successive episodes at once, and
for 9\% of them, the viewing period of a season is even shorter (1-2
days), resulting in the so-called ``binge-watching'' phenomenon. In
summary, television is no longer the main channel used to watch
\textsc{tv} series, resulting in short viewing periods of the new
seasons of \textsc{tv} series, usually released once a year.

Modern \textsc{tv} series come in various flavors: whereas
\textit{classical} \textsc{tv} series, with standalone episodes and
recurring characters, remain well-represented, they are by far not as
popular as \textsc{tv} \textit{serials}, with recurring characters
involved in a continuous plot, usually spanning several episodes, when
not several seasons: for 66\% of the people we polled, \textsc{tv}
serials are preferred to series with standalone episodes.

Yet, the narrative continuity of \textsc{tv} serials directly
conflicts with the usual viewing conditions we described: highly continuous from a
narrative point of view, \textsc{tv} serials, like any other
\textsc{tv} series, are typically watched in quite a discontinuous
way; when the episodes of a new season are released, the time elapsed
since viewing the previous season usually amounts to several months,
when not nearly one year.

As a first major consequence, viewers are likely to have forgotten to
some extent the plot of \textsc{tv} serials when they are, at last,
about to know what comes next: nearly 60\% of the people we polled
feel the need to remember the main events of the plot before viewing
the new season of a \textsc{tv} serial. Whereas discussing with
friends is a common practice to help remember the plot of the previous
seasons (used by 49\% of the people polled), the recaps available
online are also extensively used to fill such a need: before viewing a
new season, about 48\% of the people read textual synopsis, mainly in
\textit{Wikipedia}, and 43\% watch video recaps, either
``official'' or hand-made, often on \textit{YouTube}. Interestingly,
none of these ways of reducing the ``cognitive loading'' that the new
season could induce excludes the others, and people commonly use
multiple channels of information to remember the plot of \textsc{tv}
serials.

\begin{figure}[h]
  \centering
  \includegraphics[width=.9\textwidth]{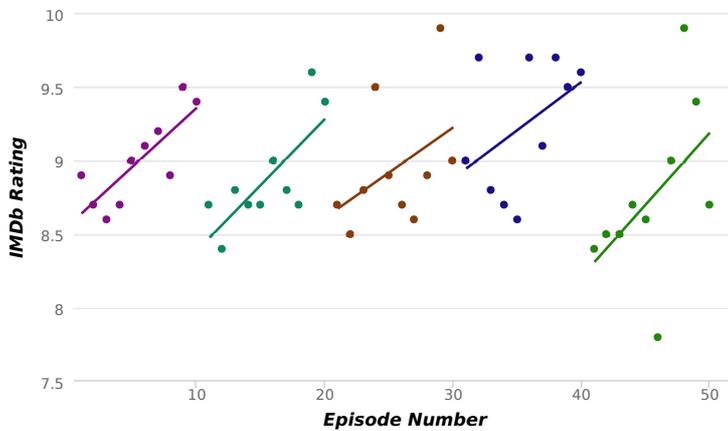}
  \caption{\label{fig:GoT_ratings}
    Average IMDb ratings of \emph{Game of Thrones}
    episodes over the first five seasons, along with season trendlines.
    }
\end{figure}

Furthermore, the time elapsed since watching the previous season may
be so long that the desire to watch the next one weakens, possibly
resulting in a disaffection for the whole \textsc{tv} serial. Many
popularity curves of \textsc{tv} serials
, as measured by their average ratings on the Internet Movie DataBase
(IMDb), can often be interpreted as exhibiting such a loss of interest
at the beginning of a new season: surprisingly, for many \textsc{tv}
serials, the average rating of each episode looks in average like a
linear function of its rank in the season, while the number of votes,
except for the very first episode of the first season, remains roughly
the same. Fig.~\ref{fig:GoT_ratings} shows such average ratings for
every episode of the first five seasons of the series \textit{Game of
  Thrones}, along with the season trendlines.

Despite intensive advertising campaigns around the new season and
possible use of cliffhangers during the previous season finale, the
season popularity trendlines tend to exhibit a kind of ``cold-start''
phenomenon, seemingly independent of the intrinsic qualities of the
season first episodes, as if the audience needed to get immersed again
in the serial universe and storylines. According to Fabrice Gobert,
the director of the French \textsc{tv} serial \textit{Les revenants},
``Writing the first episode of the first season is not an easy thing,
but the first episode of the second season is not easy either, because
we have to make the spectator want to immerse himself in the series
again.'' (radio interview on \textit{France Culture}, 09/28/2015).

In this work, we investigate a method for automatically generating
character-oriented video summaries of \textsc{tv} serials from
partially annotated data. Such character-oriented summaries are
expected to efficiently fill each user's information needs and to
tackle the cold-start issue we described by benefiting from the
empathetic relationship viewers are likely to have with some specific
characters. In order to assess our method, we performed a large scale
user study in a real-case scenario, by focusing on \textit{Game of
  Thrones} (denoted hereafter \textsc{GoT}), a popular \textsc{tv}
serial with multiple and complex storylines, a few weeks before the sixth season was publicly released.

Our main contributions are the following. The first consists in making
use of Social Network Analysis for capturing the specific dynamics of
each character's storyline. The second consists in estimating the
relevance of movie sequences in the context of summarization, by
relying to some extent on some of the stylistic patterns commonly used
by filmmakers. The third is the use of an additional criterion when
applying the standard Maximal Margin Relevance algorithm for building
the final summary. The fourth is the user study we conducted, both to
assess our method and to get valuable feedback for future work. The
last one is the annotation of the corpus that we used for experimental
purpose, which is publicly available online\footnote{\url{https://doi.org/10.6084/m9.figshare.3471839}}.

The rest of the paper is organized as follows. In
Section~\ref{sec:review}, we review the main related works. In
Section~\ref{sec:methods}, we describe the method we propose. We first
focus on the type of video units we consider as potential candidate
for later insertion in the final character-oriented summary; we then
describe the pre-processing step we perform to model the dynamics of a
specific character's storyline; and we finally detail the way we
estimate the relevance of each candidate unit, along with the
selection algorithm. In Section~\ref{sec:exp}, we describe the user
study we performed and the main results we obtained. Finally, we discuss some perspectives in Section~\ref{sec:conclu}.

\section{Related Work}
\label{sec:review}

\paragraph{TV series content-based summarization.} There is a limited
amount of work that takes narrative content into account when creating
\textsc{tv} series summaries. The most related works are
probably~\cite{Tsoneva2007}
and~\cite{Sang2010}. In~\cite{Tsoneva2007}, Tsoneva \textit{et al.}
introduce a method for automatically generating 10-minute video
summaries of standalone \textsc{tv} series episodes from the movie
scripts. In~\cite{Sang2010}, Sang \textit{et al.} introduce a way of
clustering consecutive scenes into sub-stories, before using an
attention model to build character-based summaries of full-length
movies and standalone episodes of \textsc{tv} series. However, neither of these works consider long-term narratives such as those we focus on in our work. Moreover, for \textsc{tv} serials, the method introduced in~\cite{Tsoneva2007} would result in too long summaries.

\paragraph{Plot modeling in movies and TV series.} In~\cite{Weng2009},
Weng \textit{et al.} make use of Social Network Analysis to
automatically analyze the plot of a movie: the social network
resulting from the agglomeration of every interaction between the
characters is split into communities, before narrative breakpoints are
hypothesized if the characters involved in successive scenes are
socially distant. In~\cite{Ercolessi2012}, a similar network of
interacting speakers is used, among other features, for clustering
into storylines the scenes of standalone episodes of two \textsc{tv}
series. Nonetheless, the story is considered in both of these works as only unveiling a static, pre-defined community structure within the network of interacting characters. Though such an assumption could hold for the short term plots that full-length movies and standalone \textsc{tv} series episodes depict, it is no longer the case for \textsc{tv} serials: from one episode to the other, the plot not only reveals, but also dynamically impacts the structure of the network of interacting characters. In contrast, we make in this article a dynamic use of Social Network Analysis for modeling the plot of \textsc{tv} serials. In~\cite{Tapaswi2014}, Tapaswi \textit{et al.} focus on the
interactions between the characters of standalone episodes of
\textsc{tv} series to build a visual, dynamic representation of the
plot along a timeline. Though such a representation of the narrative could generalize to \textsc{tv} serials, such a visualization focus does not provide us with tools for segmenting the plot into consistent units. In~\cite{Friedland2009}, dialogues, among other features,
are used to design a navigation tool for browsing sitcom episodes, but are considered as independent atomic events. Instead, plot analysis within \textsc{tv} serials requires us to segment the plot into larger narrative units.

\paragraph{Stylistic patterns in movies.} In~\cite{Guha2015}, Guha
\textit{et al.} adopt a style-based perspective for the plot modeling
purpose. They attempt to automatically detect the typical three-act
narrative structure of Hollywood full-length movies: each of these
three narrative segments is claimed to be characterized by specific
stylistic patterns based on film grammar; by combining low-level
features extracted from the video stream, the authors automatically
exhibit the boundaries separating these three typical consecutive
acts. In~\cite{Ma2002}, Ma \textit{et al.} introduce a video
summarization scheme based on a content-independent attention model:
some of the features used are closely related to filmmaking techniques
commonly used to make viewers focus on specific sequences, such as
shot size and music. In~\cite{Hanjalic2005}, Hanjalic \textit{et al.}
investigate low-level features, some of them based on film grammar,
like shot frequency, for modeling the emotional impact of videos, and
especially full-length movies. Such low-level features, related to
stylistic patterns used in filmmaking, are widely used in automatic
trailer generation. For instance, in~\cite{Smeaton2006}, Smeaton
\textit{et al.} make use, among other low-level features, of shot
length and camera movement to isolate action scenes for later
insertion in action movie trailers. Similarly, Chen \textit{et al.}
introduce in~\cite{Chen2004} a way of automatically generating
trailers and previews of action movies by relying on shot tempo; and
in~\cite{Smith2017}, Smith \textit{et al.} rely on an affective model
based on audio-visual features, some of them related to film grammar,
to select candidate scenes for later, manually supervised use in a
specific thriller movie trailer. However, none of these works differentiate between full-length movies
and \textsc{tv} series episodes, which are always considered as
self-sufficient from a narrative point of view. Indeed, for full-length movies or standalone episodes of classical \textsc{tv} series, stylistic patterns are probably effective enough to reliably isolate salient and meaningful sequences. Instead, we focus in this article on \textsc{tv} serials, by far the most
popular genre nowadays, defined by continuous plots and intricate narrative patterns. At this
much larger scale of dozens of episodes considered globally as
developing a single plot, possibly split into multiple, parallel
storylines, only relying on low-level stylistic features is likely to miss important developments of the narrative, semantically related to the story content; furthermore, plot modeling requires a dynamic
perspective and excludes any hypothesis about a stable and static
community structure within the network of interacting characters.

\section{Character-oriented Summaries}
\label{sec:methods}
In this section, we describe the algorithms that we use to build
character-oriented summaries of \textsc{tv} serials. We first give a general overview of our summary generation framework.

\begin{figure}[t]
  \centering
  \def\svgwidth{\columnwidth}
  \includegraphics[width=\textwidth]{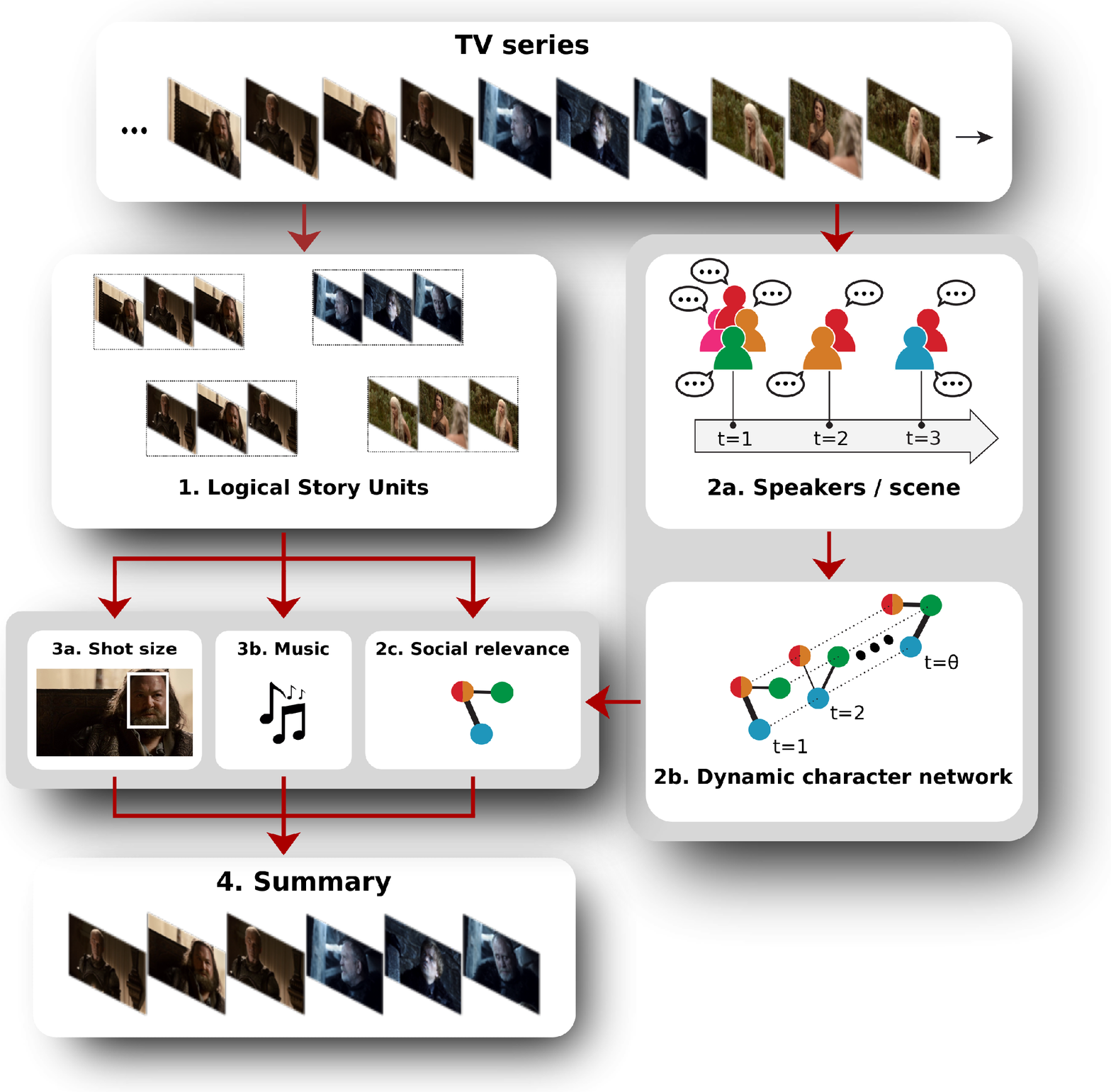}
  \caption{\label{fig:system}
    System overview. Figure available at \href{https://doi.org/10.6084/m9.figshare.7973540}{10.6084/m9.figshare.7973540} (CC-BY license).}
\end{figure}

\subsection{System Overview}
\label{subsec:system}

As can be seen on Fig.~\ref{fig:system}, the first processing step, detailed in Subsection~\ref{subsec:units}, consists in extracting from the raw video stream basic narrative units, denoted in the literature as \textit{Logical Story Units}, for potential insertion in the extractive summary. The relevance of such units is estimated according to three criteria: the first two ones, shot size (block~3a on the figure) and background music (3b) aim at capturing stylistically salient sequences, as described in Subsection~\ref{subsec:style_saliency}. The last one, social relevance (block~2c), is a content-oriented feature and aim at capturing new developments in a character's storyline. As explained in Subsection~\ref{subsec:social_relevance}, social relevance relies on the dynamic social network of interacting characters (block~2b), which in turn is based on the identification of the speakers within every scene (2a). Once estimated, shot size, background music and social relevance are combined in a single weighting scheme, and relevant Logical Story Units are iteratively selected according to the alg;orithm detailed in Subsection~\ref{subsec:summ}, resulting in the final summary shown on Fig~\ref{fig:system} (block~4).

\subsection{Logical Story Unit Detection}
\label{subsec:units}

In this subsection, we define the video sequences we regard as the
basic candidate units for potential, later insertion in the summary,
along with a novel algorithm for extracting them.

Usually built upon sophisticated editing rules, the ``official'' video
recaps of \textsc{tv} serials rarely concatenate single shots
extracted from different parts of the original stream. Instead, the
basic unit used in such summaries is typically a short sequence of
about 10 seconds consisting of a few consecutive shots. Such sequences
are usually selected not only because of their semantic relevance, but
also because of their semantic cohesion and self-sufficiency. From a
computational perspective, identifying such sequences in the video
stream as potential candidates for later insertion in the final
summary remains tricky.

The stylistic patterns widespread among filmmakers are particularly
relevant, because they are often used to emphasize the semantic
consistency of these sequences. For instance, dialogue scenes require
the ``180-degree'' rule to be respected so as to keep the
exchange natural enough: in order for both speakers to seem to look at
each other when they appear successively on-screen, the first one
must look right and the second one must look left. To achieve this,
two cameras must be placed along the same side of an imaginary line
connecting them. Such a rule results in a specific visual pattern made
of two alternating, recurring shots and is highly typical of dialogue
scenes.

\begin{figure}[h]
  \centering
  \includegraphics[width=\textwidth]{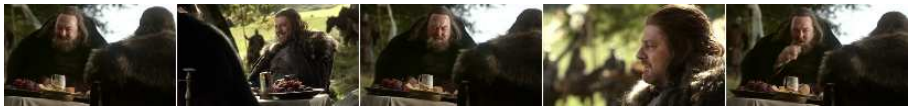}
  \caption{\label{fig:lsu} Example of Logical Story Unit (LSU) with
    one surrounding, recurring shot in positions 1, 3, 5.}
\end{figure}

More generally, several sets of such recurring shots may overlap each
over, resulting in possibly complex patterns well-suited for
segmenting movies into consistent narrative
episodes. In~\cite{Hanjalic1999}, Hanjalic \textit{et al.} denote as \textit{Logical Story Units} (\textsc{lsu}s) such sequences
of intertwined recurring shots, and introduce a method for automatically
extracting them. Fig.~\ref{fig:lsu} shows a sequence of five shots
with one recurring shot in positions 1, 3, 5, resulting in one
\textsc{lsu}.

We use \textsc{lsu}s as the basic candidate units selected when building
the summaries of \textsc{tv} serials. In order to detect such
\textsc{lsu}s, we first split the whole video into shots, which we
then compare and label according to their similarities. Both tasks,
shot cut detection as well as shot similarity detection, rely on image
comparison: we first model images with $3$-dimensional histograms of the image pixel values in the \textsc{hsv} color space, before performing the standard block-based comparison technique detailed in~\cite{Koprinska2001} by Koprinska \textit{et al.}
Once identified, similar shots can be used as a basis for automatically
extracting every \textsc{lsu}.

Instead of the standard graph-based algorithm Yeung \textit{et al.} describe in~\cite{Yeung1998}, we introduce here a novel, alternative matrix-based
algorithm to automatically detect \textsc{lsu} boundaries. Though
computationally more expensive, such an algorithm has the advantage of
being more straightforward to implement. The resulting sequences are
strictly the same as when applying the standard algorithm: both approaches only
depend on the reliability on the previous shot similarity detection
step, which performed pretty well when applied to a subset of
annotated episodes ($F$-score $\simeq 0.90$).

Once performed, shot similarity detection results in a symmetric
similarity matrix $\mathcal{S}$, where $s_{i,j}$ is set to $1$ if the
$i$\textsuperscript{th} and $j$\textsuperscript{th} shots are
considered as similar, and to $0$ otherwise. As long as the shots
are chronologically ordered, such a representation constitutes a
straightforward way of automatically detecting the \textsc{lsu}
boundaries.

For the five shots included in the sequence shown on
Fig.~\ref{fig:lsu}, the similarity matrix $\mathcal{S}$ is filled as
follows:

\begin{equation}
  \mathcal{S} =
  \begin{pmatrix}
    1 & 0 & 1 & 0 & 1 \\
    0 & 1 & 0 & 0 & 0 \\
    1 & 0 & \tikzmark{left1}1\tikzmark{right1} & 0 & 1 \\
    \tikzmark{left2}0 & 0 & 0 & \tikzmark{left3}1\tikzmark{right3} & 0 \\
    \tikzmark{left4}1 & 0\tikzmark{right2} & 1\tikzmark{right4} & 0 & 1
  \end{pmatrix}
  \label{eq:lsu_mat}
\end{equation}

\DrawBox[thick, blue]{left1}{right1}{\textcolor{blue}{}}
\DrawBox[thick, blue, dashed]{left2}{right2}{\textcolor{blue}{}}
\DrawBox[thick, red]{left3}{right3}{\textcolor{red}{}}
\DrawBox[thick, red, dashed]{left4}{right4}{\textcolor{red}{\footnotesize$S^{(4)}$}}

\medskip

The $k$\textsuperscript{th} shot ($1 < k < n$, where $n$ is the total
number of shots) is strictly included in one \textsc{lsu} if
surrounded by at least two occurrences of the same recurring shot: in
the matrix $\mathcal{S}$, such a statement is equivalent to the fact that
the double sum $S^{(k)} := \sum_{(i > k, j < k)} s_{i,j}$ is greater
or equal to $1$.  In Equation~\ref{eq:lsu_mat}, the terms of the sum
$S^{(4)}$ are included in the dashed red box. The fact that $S^{(4)}
\geqslant 1$ means that the 4\textsuperscript{th} shot (solid red box)
is surrounded by at least two occurrences of the same recurring shot,
the first one occurring after ($i > 4$) and the second one before ($j
< 4$) the 4\textsuperscript{th} position, and strictly belongs to one
\textsc{lsu}.

The \textsc{lsu} boundaries can then be deduced from the two
quantities $S^{(k)}$ and $S^{(k-1)}$, ($1 < k < n$, with $S^{(1)} :=
0$) according to the two following rules: 1) if $S^{(k-1)} = 0$ and
$S^{(k)} \geqslant 1$, the $(k-1)$\textsuperscript{th} shot is the
beginning of a new \textsc{lsu}; and 2) conversely, if $S^{(k-1)}
\geqslant 1 $ and $S^{(k)} = 0$, the $k$\textsuperscript{th} shot is
the end of the previous \textsc{lsu}.  Furthermore, the double sum
$S^{(k)} := \sum_{(i > k, j < k)} s_{i,j}$ does not need to be
computed for each $k = 2, ..., (n-1)$ but can be recursively deduced
from the previous quantity $S^{(k-1)}$ according to the following
relation:

\begin{equation}
  S^{(k)} = S^{(k-1)} - \sum_{j < k - 1} s_{k, j} + \sum_{i > k} s_{i, k - 1}
\end{equation}

In the example of Equation~\ref{eq:lsu_mat}, the quantity $S^{(4)}$ (sum of the
coefficients inside the dashed red box) can then be recursively
obtained from the quantity $S^{(3)}$ (sum of the coefficients
inside the dashed blue box) as follows: $S^{(4)} = S^{(3)} - (s_{4,1}
+ s_{4,2}) + (s_{5,3})$.

By construction, the value of the coefficient $s_{k, k-1}$ is equal to
$0$ (two consecutive shots cannot be the same) and is ignored when
recursively updating the quantity $S^{(k)}$ from $S^{(k-1)}$.

The method we use requires two nested loops over every shot,
resulting in a time complexity in $O(n^2)$. Nonetheless, such a method
captures maximal \textsc{lsu}s, often far too long to be inserted into
a summary of acceptable length.



In order to get shorter candidate sequences without loosing the formal
consistency of the \textsc{lsu}s, we apply recursively
our extraction algorithm within each \textsc{lsu} to obtain more
elementary, not maximal, \textsc{lsu}s. During the extraction process,
we put both lower and upper bounds on the duration of the candidate
\textsc{lsu}s (either maximal or more elementary). Based on what we observe in the manually edited official recaps of \textsc{tv} serials, we constrained
every final candidate \textsc{lsu}s to last at least $5$ seconds, and
at most $15$.

We then estimate the relevance of each precomputed \textsc{lsu} for
possible insertion into the summary according to three criteria. The first
criterion is related to the content of the storyline associated to the
considered character. The two other ones rely on techniques commonly
used by filmmakers to tell the story, and are related to the form of
the narrative.

\subsection{Social Relevance}
\label{subsec:social_relevance}

\paragraph{Narrative episode.} In order to estimate the relevance of each
\textsc{lsu} content, we first automatically segment the storyline
associated to a specific character into \textit{narrative episodes}. In any narrative, the story of a specific character usually develops
sequentially and advances in stages: each narrative episode is defined
as a homogeneous sequence where some event directly impacts a specific
group of characters located in the same place at the same time. Though
such a notion of narrative episode may be defined at different levels
of granularity, such sequences are often larger than the formal
divisions of books in chapters and of \textsc{tv} serials in
episodes. Here are some examples of character-based narrative episodes
in \textit{Game of Thrones}: ``Theon Greyjoy rules Winterfell'';
``Arya Stark captive in Harrenhal''; ``Jaime Lannister's journey in
Dorne to rescue Myrcella''...

The segmentation of each character's storyline into narrative
episodes aims at building a summary able to capture the dynamics of
the plot and is performed as follows.

We first pre-compute the weighted,
undirected dynamic social network of interacting characters over the whole \textsc{tv} serial according to the method introduced in~\cite{Bost2016a} and detailed in~\cite{Bost2018}. Such a dynamic network is built upon the speaker turns and scene boundaries and is based on a smoothing method that provides us with an instantaneous view of the state of any relationship at any point of the story, whether the related characters are interacting or not at this moment. As a result, the full, smoothed, social neighborhood of a specific character is always available at any time $t$.

As stated above, building the dynamic social network of interacting characters heavily depends on two key steps: scene boundaries detection and speaker detection. Though it would have been possible to automatically perform both tasks, the second one (speaker detection) either in a supervised (speaker recognition) or in an unsupervised way (speaker diarization), we decided to hand-label the data. First,
we wanted to do a relatively large scale user study and could not afford to
show the viewers in a limited time both fully and partially
automatic summaries to measure the impact of the errors made at the
speaker recognition/diarization level. Second, \textsc{tv} serials usually contain
many speakers, even when focusing on the major ones, often speaking in
adverse conditions (background music, sound effects) resulting, as reported in~\cite{Bredin2016}, in high diarization error rates. We left the task of speaker recognition/diarization in such challenging conditions
for future work.

Based on the dynamic network of interacting characters, we define $\mathbf{r_t}$ as the relationship
vector of a specific character at time $t$: $\mathbf{r_t}$ contains
the weights, ranging between $0$ and $1$, of his/her relationships
with any other character at time $t$. Here are two examples of such
relationship vectors for the \textit{Game of Thrones} character
\textit{Arya Stark}, respectively in the 34\textsuperscript{th} and
49\textsuperscript{th} scene where she appears (the components are
re-arranged in decreasing order of importance):

\begin{equation}
\mathbf{r_{34}} =
\left (
  \begin{array}{ll}
    \text{Tywin} & [0.82] \\
    \text{Jaqen} & [0.23] \\
    \text{Hot Pie} & [0.21] \\
    \text{A. Lorch} & [0.21] \\
    \multicolumn{2}{c}{\vdots}
  \end{array}
  \right )
  \
  \mathbf{r_{49}} =
\left (
  \begin{array}{ll}
    \text{Beric} & [0.54] \\
    \text{Thoros} & [0.51] \\
    \text{Anguy} & [0.51] \\
    \text{Clegane} & [0.50] \\
    \multicolumn{2}{c}{\vdots}
  \end{array}
  \right )
  \label{eq:relationships}
\end{equation}

For each specific character we target, we then compute the distance matrix
$\mathcal{D}$, where $d_{t, t'}$ is the normalized Euclidean distance
between the character's relationship vectors $\mathbf{r_t}$ and
$\mathbf{r_{t'}}$ at times $t$ and $t'$. Because each narrative
episode is defined as impacting a limited and well-identified group of
interacting characters, the relationships of a character are expected
to stabilize during each narrative episode, and to change whenever a new
one occurs.

\begin{figure}[h]
  \centering
  \includegraphics[width=.65\textwidth]{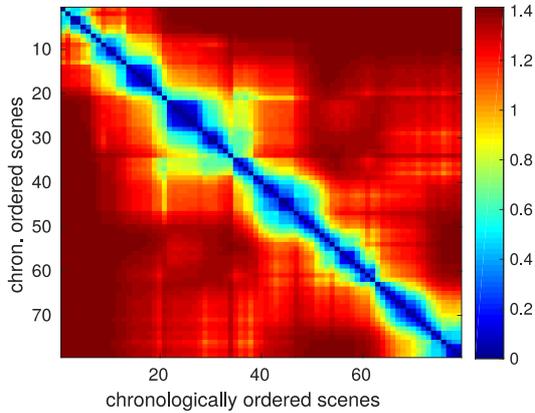}
  \caption{\label{fig:arya_neighbors} Matrix of distances between Arya
    Stark's relationships in any pair of scenes $(t, t')$ in
    the first five seasons of \textsc{GoT}. Figure available at \href{https://doi.org/10.6084/m9.figshare.7973540}{10.6084/m9.figshare.7973540} (CC-BY license).}
\end{figure}

Fig.~\ref{fig:arya_neighbors} shows the matrix $\mathcal{D}$ for the
character \textit{Arya Stark}, as built over the first five seasons of
\textsc{GoT}. In this matrix, the time steps are the scenes, ordered
chronologically, and for the sake of clarity, we build the matrix only
upon the scenes where the character is involved, even though the
smoothing method we described in~\cite{Bost2018} provides a way of
estimating the social neighborhood of the character in any scene.

As can be seen from Fig.~\ref{fig:arya_neighbors}, the character's
social environment is not continuously renewed as the storyline
develops, but stabilizes for some time, before being replaced by a new
social configuration. For instance, between scenes 38 and 51, the
social environment of Arya remains quite the same, suggesting that her
storyline stabilizes in some narrative episode. Interestingly, other
narrative episodes can also be observed in the matrix at larger
(scenes 6--48) or smaller (scenes 21--26) scales, confirming the
relative and multi-scale nature of the notion of narrative episode.

\paragraph{Optimal partitioning.} We then optimally partition the
distance matrix $\mathcal{D}$, so that to split the whole character's
storyline into successive narrative episodes. Such a partitioning
depends on a threshold $\tau$ set by the user himself, depending on
his specific information needs: $\tau$ corresponds to
the maximal admissible distance between the most covering relationship
vector in each narrative episode and any other relationship vector
within this narrative episode; it can be interpreted as the level of
granularity desired when analyzing the story.

We partition the whole set of scenes (storyline) into disjoint subsets
of contiguous scenes (narrative episodes) by adapting a standard set
covering problem to this partitioning purpose (see for
example~\cite{Daskin2011} for the standard formulation of the
set covering problem). First, a constraint of temporal contiguity is
put on the elements of the admissible subsets of scenes, so that to
keep narrative episodes continuous over time. Second, in order to
obtain a covering as close as possible to a partition, we minimize the
overlapping between the covering subsets instead of minimizing their
number as in the standard formulation of the set covering
problem. Despite this adapted objective, some relationship vectors at
the boundaries between two consecutive narrative episodes may still
belong to both of them: in this case, the covering is refined into a
real partition by assigning the duplicated vector to the closest
relationship state.

\begin{figure}[h]
  \centering
  \includegraphics[width=.65\textwidth]{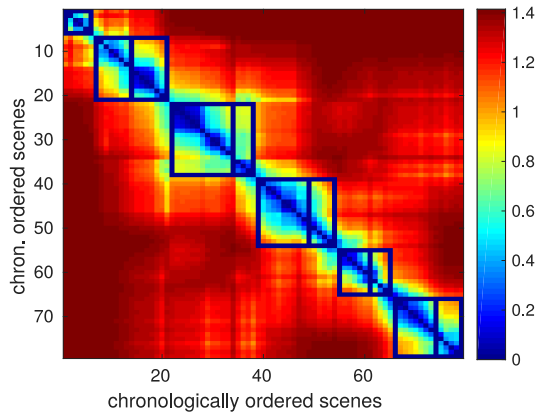}
  \caption{\label{fig:arya_neighbors_part}Partitioned matrix of
    distances between Arya Stark's relationships in any pair
    of scenes $(t, t')$ in the first five seasons of \textit{Game of
      Thrones}. Figure available at \href{https://doi.org/10.6084/m9.figshare.7973540}{10.6084/m9.figshare.7973540} (CC-BY license).}
\end{figure}

Fig.~\ref{fig:arya_neighbors_part} shows the resulting partition of
Arya Stark's distance matrix $\mathcal{D}$
(Fig.~\ref{fig:arya_neighbors}), for a granularity level $\tau =
1.0$. Each narrative episode is represented as a box containing a
vertical line. This line corresponds to the scene in which the
relationship state covers at best the narrative episode. In such
scenes, the relationship vector can be regarded as conveying the
typical social environment of the character within the associated
narrative episodes. 

The two relationship vectors $\mathbf{r_{34}}$ and $\mathbf{r_{49}}$
that we introduced in Equation~\ref{eq:relationships} (3\textsuperscript{rd} and 4\textsuperscript{th}
vertical lines on Fig.~\ref{fig:arya_neighbors_part}) best cover
Arya Stark's social neighborhood in the third and fourth narrative
episodes. These two relationship vectors turn out to perfectly match
two major developments in Arya's story: ``Arya Stark captive in
Harrenhal'' ($\mathbf{r_{34}}$) and ``Arya Stark and the Brotherhood''
($\mathbf{r_{49}}$).

\paragraph{Social relevance.} For the considered character, the
\textit{social relevance} $\mathbf{sr_i}$ of the
$i$\textsuperscript{th} \textsc{lsu} is then defined as the cosine
similarity between the representative vector $\mathbf{r_t}$ of the
character's relationships in the narrative episode to which the
$i$\textsuperscript{th} \textsc{lsu} belongs and the vector of
relationships the character is currently having within the
$i$\textsuperscript{th} \textsc{lsu}. As mentioned before, the
representative vector $\mathbf{r_t}$ is derived from the smoothing
method detailed in~\cite{Bost2018}, however the components of the
character's relationships vector within each \textsc{lsu} correspond
to the interaction times between the character and every other
character. The interaction time is estimated according to the basic
heuristics described in~\cite{Bost2018}. For a specific character,
such a \textit{social relevance} measure aims at discriminating the
\textsc{lsu}s showing some of his/her typical relationships within each
narrative episode of his storyline. Nonetheless, social relevance
remains too broad a criterion to be used on its own for isolating
relevant video sequences when building the summary. In the next
subsection, we focus on two additional, stylistic features that can
help to isolate salient \textsc{lsu}s among all those that are equally
relevant from the social point of view.

\subsection{Stylistic Saliency}
\label{subsec:style_saliency}

The semantics of most movie sequences depends not only on their
objective contents, but also on the way they are filmed and
edited. For instance, the importance of a specific sequence in the
plot, though primarily dependent on the content of the associated
event, is usually emphasized by some stylistic patterns commonly used
in filmmaking. We here focus on two such stylistic patterns to isolate
salient sequences among all the possible \textsc{lsu}s: on the one
hand the size of the shots and on the other the background music. We
selected both of them for their reliability when isolating salient
video sequences in movies, and because of their low computational
cost.

\begin{figure}[h]
  \centering
  \includegraphics[width=.8\textwidth]{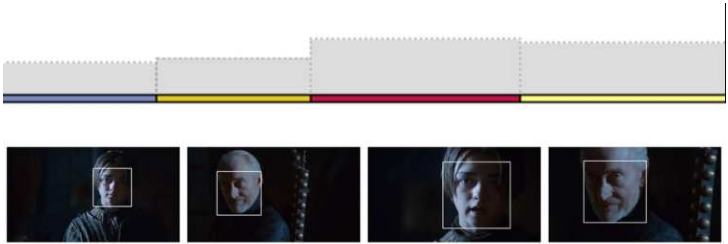}
  \caption{\label{fig:face_seq} Shot sequence with face boundaries as
    automatically detected. On the top part, the height of the gray
    rectangles corresponds to the shot size, as a proportion of the
    frame height.}
\end{figure}

\paragraph{Shot size.} The shot sizes are estimated by applying the face
detector described in~\cite{Felzenszwalb2008} to a
sample of 5 video frames for each shot. The 5 frames are uniformly
distributed over time within the shot. Such a sample size of 5 frames
aims both at keeping the computation time reasonable when performing
face detection, and at facing the issue of the characters adopting
various poses during a single shot, which is likely to cause false
negatives.  For each of the 5 frames, we retain, if any, the largest
face box. We then pre-compute the shot size as the median height of all
the face boxes detected over the 5 frames and we express it as a
proportion of the video frame height. Using the median rather than the
mean value prevents the shot size from being biased by very large or
very small false positives, usually hypothesized for a single frame
only.  Fig.~\ref{fig:face_seq} shows a sequence of four shots along
with the face boundaries as automatically detected; on the top of the
figure, the height of the gray rectangles corresponds to each shot
size, as a proportion of the frame height.  The shot size
$\mathbf{ss_i}$ we obtain for each $i$\textsuperscript{th}
\textsc{lsu} is the mean value of the size of the shots it contains,
resulting, besides the social relevance $\mathbf{sr_i}$, in a second,
style-oriented, feature for possible insertion into the final summary.

\paragraph{Musicality}. The second stylistic feature we consider is
music. Based on the \textsc{matlab} \textit{MIRtoolbox} package
described in~\cite{Lartillot2008}, we implemented a basic music
tracker relying on the method Giannakopoulos \textit{et al.}
introduced in~\cite{Giannakopoulos2008} for tracking music in
movies. The features we use to distinguish the background music from
speech rely on the chroma vectors, conveying the distribution of the
audio signal over the twelve notes of the octave. For music, the audio
signal typically results in chroma vectors with components both less
uniformly distributed over the octave and more stable over time than
for speech.

The musicality $\mathbf{m_i}$ of the $i$\textsuperscript{th} Logical
Story Unit is then pre-computed according to the method described
in~\cite{Giannakopoulos2008} to capture the statistical dispersion of
the chroma vector, both over the twelve notes and over time, resulting
in a single scalar feature indicative of the average musicality of each
\textsc{lsu}.

\subsection{Selection Algorithm}
\label{subsec:extraction}

The three features we introduced (social relevance, average shot size and average musicality) are combined for each \textsc{lsu} into a single measure of relevance $p_i$ estimated according to the following weighting scheme:

\begin{equation}
	p_i = \lambda_1 \mathbf{sr_i} + \lambda_2 \mathbf{ss_i} + \lambda_3 \mathbf{m_i}
    \label{eq:relevance}
\end{equation}

By construction, social relevance and average shot size range from
$0$ to $1$. We therefore min-max normalize the average musicality to
get values between $0$ and $1$. In the use case that we detailed in Section~\ref{sec:intro}, the respective weights of these three features would be set by the users themselves, depending on their specific needs: on the one hand, emphasizing social relevance is expected to result in more informative summaries, able to help the user remember the plot; on the other hand, emphasizing music and shot size is expected to result in trailer-like summaries, able to address the cold-start phenomenon we described in Section~\ref{sec:intro}. We defer to Subsection~\ref{subsec:summ} the discussion of the particular weight settings we used in our contrastive experimental study. 

In the rest of this subsection, we describe how we build the summary
of the storyline associated to a specific character by iteratively
selecting optimal candidate \textsc{lsu}s, once their
relevance and duration are set.

Character-oriented summaries aim at reflecting the dynamics of a
character's storyline. Once isolated by applying the segmentation
method described in Subsection~\ref{subsec:social_relevance}, each
narrative episode of a specific character's storyline should be
equally reflected, whatever its duration, as a major development in
his/her story. We therefore build the summary step-by-step to reflect the
natural segmentation of the storyline into narrative episodes,
possibly of variable duration.

Besides the weighting scheme of Equation~\ref{eq:relevance}, our algorithm for constructing the character-oriented summary takes two inputs set by the users themselves, depending on their information needs: the level of granularity for analyzing the storyline in narrative episodes; and the maximum time $T$, expressed
in seconds, devoted in the summary to each narrative episode.

Each narrative episode, once isolated, consists of a subset of scenes,
containing $n$ candidate \textsc{lsu}s, each $i$\textsuperscript{th}
\textsc{lsu} being weighted according to the global relevance score
$p_i$ introduced in Equation~\ref{eq:relevance}. Summarizing the narrative episode can then be regarded as a task with two joint objectives and a
length constraint: the summary must not exceed the duration $T$ and
aims at containing not only relevant sequences, but also sequences
that remain as diverse as possible, in order to minimize
redundancy.

In~\cite{Mcdonald2007}, McDonald shows that such a summarization
problem can be formulated as the following quadratic knapsack problem,
with two joint objectives and a length constraint:

\begin{equation}
  \text{(P)}
  \left \{
  \begin{aligned}
    & \max f(\mathbf{x}) = \left ( \sum_{i=1}^n p_i x_i + \sum_{i=1}^n \sum_{j=1}^n d_{ij} x_i x_j \right ) \\
    \mathrm{s.t.} 
    & \left \bracevert
    \begin{aligned}
      & \sum_{i = 1}^n w_i x_i \leqslant T & i = 1, ..., n \\
      & x_i \in \{0, 1\} & i = 1, ..., n \\
    \end{aligned}
    \right. \\
  \end{aligned}
  \right.
  \label{eq:summ_prob}
\end{equation}

\noindent where $x_i$ is a binary variable set to $1$ if the
$i$\textsuperscript{th} \textsc{lsu} is inserted in the narrative episode summary, and
to $0$ otherwise; $p_i$ denotes the relevance, as computed according
to Equation~\ref{eq:relevance}, of the $i$\textsuperscript{th}
\textsc{lsu}; $w_i$ is the duration, expressed in seconds, of the
$i$\textsuperscript{th} \textsc{lsu}; $T$ is the maximum time devoted
in the summary to the narrative episode containing the current subset
of $n$ \textsc{lsu}s; and finally, $d_{ij}$ is a measure of dissimilarity
between the $i$\textsuperscript{th} and $j$\textsuperscript{th}
\textsc{lsu}s: it is defined as the normalized Euclidean distance
between the vectors of relationships in the $i$\textsuperscript{th}
and $j$\textsuperscript{th} \textsc{lsu}s, as defined in
Subsection~\ref{subsec:social_relevance}.

Once introduced in the objective function, such a coefficient $d_{ij}$
aims at generating a summary that provides us with an overview of the
full range of the character's relationships at this point of the
story, instead of focusing on a single relationship shown in several
redundant sequences.

As stated in~\cite{Mcdonald2007} and in~\cite{Gillick2009}, such a
formulation of the summarization problem can be tricky to solve
exactly for large instances, even when linearizing the quadratic part
of the objective function, and heuristic methods provide us with much more scalable, though possibly sub-optimal, resolution
techniques. We introduce here a greedy heuristic for iteratively
selecting optimal \textsc{lsu}s. In~\cite{Mcdonald2007}, McDonald
underlines the limitations of Maximal Margin Relevance-based
algorithms (\textsc{mmr}) to reach the double objective of relevance
and diversity when building summaries: if selected iteratively without
taking their length into account, the already selected sequences may
be too long and prevent us from choosing additional sequences to
improve the objective function.

As a possible alternative to the dynamic programming-based algorithm
detailed in~\cite{Mcdonald2007}, we introduce here the method summarized
in Algorithm~\ref{algo:select_lsu}. It generalizes to the quadratic
case the usual greedy heuristic used to solve the linear knapsack
problem in~\cite{Fayard1982} and~\cite{Balas1980}.

\begin{figure}[h]
  \centering
  \begin{algorithmic}[1]
    \REQUIRE $\mathbf{p}, \mathbf{w} \in \mathbb{R}_+^n, \ T \in \mathbb{R}_+, (d_{ij})_{i, j = 1, ...n} \in \mathbb{R}_+$
    \STATE $L \leftarrow \{1, ..., n\}$
    \STATE $S \leftarrow \emptyset$
    \WHILE {$L \neq \emptyset$ \AND $T > 0$}
    \STATE $s \leftarrow \argmax_{i \in L} (p_i / w_i)$ \label{algo:ratio}
    \IF {$w_s \leqslant T$ \AND $\mathrm{non\_overlap}(s, S)$} \label{algo:noOverlap}
    \STATE $S \leftarrow S \cup \{s\}$
    \STATE $T \leftarrow T - w_s$
    \FOR {$i \leftarrow 1$ to $n$}
    \STATE $p_i \leftarrow p_i + 2d_{is}$ \label{algo:updtRelevance}
    \ENDFOR
    \ENDIF
    \STATE $L \leftarrow L - \{s\}$
    \ENDWHILE
    \RETURN $S$
  \end{algorithmic}
  \caption{\label{algo:select_lsu} \textsc{LSU} selection algorithm}
\end{figure}

The summary $S$, containing the indices of the selected \textsc{lsu}s,
is built iteratively from the whole subset $L$ of candidate
\textsc{lsu}s: at each iteration, we choose the \textsc{lsu} with the
maximum relevance/duration ratio (line~\ref{algo:ratio}) that
simultaneously does not exceed the total duration limit and does not
overlap with any of the previously selected \textsc{lsu}s (conditions
line~\ref{algo:noOverlap}): some candidate \textsc{lsu}s may share
some shots and partially overlap.  Such a criterion tends to select
short \textsc{lsu}s, as long as they are both relevant and
diverse. The objective of diversity is taken into account when
iteratively updating the vector of relevance values
(line~\ref{algo:updtRelevance}): the updating formula comes from the
re-writing of the objective function after each new sequence is
selected. After the very first iteration of the algorithm for example,
and assuming, without loss of generality, that the first \textsc{lsu}
is selected, the objective function can be re-written as:

\[
f(\mathbf{x}) = (p_1 + d_{11}) + \left ( \sum_{i=2}^n (p_i + 2d_{i1}) x_i + \sum_{i=2}^n \sum_{j=2}^n d_{ij} x_i x_j \right )
\]

\noindent where the right-hand part of the equation is the objective
function of the summarization problem \ref{eq:summ_prob} for the
remaining, not yet selected, \textsc{lsu}s, again formulated as the
objective function of a quadratic knapsack problem, but with an
updated vector of relevance values. As reported in Subsection~\ref{subsec:summ}, Algorithm~\ref{algo:select_lsu}, with a time complexity in $O(n^2)$, results in short computation times, even for large instances of the problem. Moreover, the provided solution turn out to be very close to the optimal solution, but much more scalable.

For each narrative episode, a
summary is built until the time duration limit $T$ is reached, and the
final character-oriented summary is made of the concatenation of all
the \textsc{lsu}s, chronologically re-ordered, selected in every
narrative episode.

\section{Experiments and Results}
\label{sec:exp}

In order to assess our method for automatically generating
character-oriented summaries of \textsc{tv} serials, we performed a
large scale user study in a real case scenario. In this section, we
first describe the user study we performed, we then explain the types
of summaries the participants were asked to rank, along with the
evaluation protocol, and we finally detail the obtained results.

\subsection{User Study}
\label{subsec:user_study}

A few weeks before the sixth season of the popular \textsc{tv} serial
\textit{Game of Thrones} was released, people, mainly students and
staff of our university, were asked to answer a questionnaire, both in
order to collect various data about their \textsc{tv} series viewing
habits and to evaluate automatic summaries centered on five characters
of \textsc{GoT}. The responses are reported in full detail
in~\cite{Bost2016b}, Appendix~A. A total of 187 subjects took part in
the questionnaire, with 52.7\% female, and 47.3\% male
participants. The population was quite young: 21.14 years old in
average, with a standard deviation of 2.67 years. Being familiar with
\textsc{GoT}, if recommended, was not mandatory to answer the
questionnaire. 27\% of the people we polled had actually never watched
\textsc{GoT} when answering the questionnaire, but 56\% of them had
seen all first five seasons.  More than half of the \textsc{GoT}'s
viewers we polled in our study stated that they feel the need to
remember the plot of the past seasons when a new one is about to be
released. To do so, they use multiple channels of information:
discussions with friends (57.6\%), textual (32.9\%) and video
summaries (34.1\%).


Among the people who watched every past season of \textsc{GoT}, 64.2\%
had last watched an episode of this \textsc{tv} serial more than six
months ago, and were in the typical use case we described in
Section~\ref{sec:intro}.

\subsection{Summaries for Evaluation}
\label{subsec:summ}

After answering general questions about their \textsc{tv} series
viewing habits, the participants were asked to evaluate summaries of
\textsc{GoT} centered on the storylines of five different characters
to ensure generalizability: Arya Stark, Daenerys Targaryen, Jaime
Lannister, Sansa Stark and Theon Greyjoy.

\paragraph{TV series data.} By focusing on such a popular \textsc{tv}
serial, we were quite certain that most participants would have
watched it; moreover, with \textsc{GoT}, we were at the time of the study in the typical use
case that we want to target (see Section~\ref{sec:intro}), with a new
season about to be released.

The dataset covers the whole set of the first five seasons (50 one-hour
episodes) of \textsc{GoT}. As stated in Subsection~\ref{subsec:social_relevance}, the summaries are generated from partially annotated data: we manually inserted the scene boundaries within each
episode and we labeled every subtitle according to the corresponding
speaker, as a basis for estimating verbal interactions between
characters and building the dynamic social network of interacting
characters. Every other step (\textsc{lsu} detection, shot size and musicality estimation, summary generation) is performed automatically.

\paragraph{Summary generation.} Though in a real system characters would be selected on-demand by the users themselves, we had to focus on specific ones for assessing our approach in a common setting. Our criteria to select the five characters were the following: these characters were still
involved in the narrative at the end of the fifth season of
\textsc{GoT}, and were therefore likely to play a role in the next one;
they all are important enough to have evolved at some point of the
plot in their own storylines; their story is likely to be complex
enough to require a summary before viewing the next season.

The summaries cover the storylines of the five considered characters
over all the first five seasons of \textsc{GoT}. Such long-term
summaries were expected to capture the whole dynamics of a character's
storyline when introducing to the next season, rather than only
focusing on the very last events he happened to experience during the
last season. The more a plot is advanced, the more such long-term
summaries are probably needed, especially when the plot is complex.

We automatically generated three summaries for each character:

First, a full summary (denoted \textbf{full}), built upon the method
we described in Section~\ref{sec:methods} and designed so as to be
sensitive to both the content and the style of the narrative. This
first summary depends on three parameters: the vector $\bm{\lambda}$ of feature weights introduced in Equation~\ref{eq:relevance} to estimate the relevance of each \textsc{lsu}; 
 the granularity level $\tau$ used for segmenting the storyline into
narrative episodes; and the time $T$ devoted in the summary to each of
them. In a real system, these three parameters would be set by the users themselves, depending on their specific needs. In contrast, in our user study, we chose a particular parameter setting to ensure methodologically sound comparison: in order to keep the summary duration into reasonable boundaries, we set the granularity level to $\tau = 1.0$ and the duration devoted to each narrative episode to $T = 25$ seconds; for generating such full summaries, we set as explained below the feature weights to $\bm{\lambda} := (0.16, 0.42, 0.42)^\intercal$.

A second, style-based summary (denoted \textbf{sty}), is built only upon
the stylistic features we described in
Subsection~\ref{subsec:style_saliency}, by setting the feature weights to $\bm{\lambda} := (0, 0.5, 0.5)^\intercal$. By discarding social relevance, there is no longer need for the
pre-segmentation of the storyline into narrative episodes. As a consequence, the candidate
\textsc{lsu}s are not selected among the separate subsets resulting
from the segmentation step; instead they are considered as a whole
single set of candidate sequences, weighted equally according to their average
musicality and shot size, and finally selected by iteratively applying Algorithm~\ref{algo:select_lsu} until the resulting style-based
summary has roughly the same duration as the full summary.

The specific feature weight values $\bm{\lambda} := (0.16, 0.42, 0.42)^\intercal$ we chose for building the full summaries were set as follows: we increased the weight of social relevance in the weighting scheme of Equation~\ref{eq:relevance} (from 0 to 0.16) until the resulting full summaries differed significantly (by $\simeq 66\%$) from the style-oriented ones. Such a contrastive methodology was expected to measure the benefits of incorporating social relevance for capturing the plot content, in addition to the equally weighted style-oriented features (music and shot size). 

Third, a baseline, semi-random summary (denoted \textbf{bsl}) is
obtained as follows: some non-overlapping \textsc{lsu}s where the
considered character is verbally active are first randomly selected
until reaching a duration comparable to the duration of the first two
kinds of summaries; the selected \textsc{lsu}s are then re-ordered
chronologically when inserted in the summary.

The main properties of the resulting three types of summaries are
reported in Table~\ref{tab:summ_features} for each of the five
characters considered. For each character and each type of summary, the
number of candidate \textsc{lsu}s is mentioned, along with their
average duration in seconds. The same properties are reported for the
selected \textsc{lsu}s inserted in the summary. The duration of the
resulting summary, expressed in seconds, is mentioned in the seventh
column. The compression rate $r$ is mentioned in the eighth
one, and is expressed as the ratio between the total duration of all the scenes
in which the character is verbally active and the summary duration. Finally, the computation time, expressed in seconds, needed for generating each summary is reported in the ninth, last column.  

\begin{table}[h]
  \renewcommand{\arraystretch}{1.0}
  \caption{\label{tab:summ_features} Properties of the three types of summary generated for each character's storyline during the first five seasons of \textsc{GoT}: number and average duration of candidate and selected \textsc{lsu}s, summary duration and compression rate.}
  \centering
  \begin{tabular}{|c|c|rr|rr|r|c|c|}
    \hline
    \multirow{3}*{Character} & \multirow{3}*{Summary} & \multicolumn{4}{c|}{\textsc{lsu}s} & \multirow{3}*{Dur.} & \multirow{3}*{$r$} & \multirow{3}*{Gen. t.} \\
    \cline{3-6}
    & & \multicolumn{2}{c|}{Candidate} & \multicolumn{2}{c|}{Selected} & & & \\
    \cline{3-6}
    & & \multicolumn{1}{c}{\#} & \multicolumn{1}{c|}{dur.} & \multicolumn{1}{c}{\#} & \multicolumn{1}{c|}{dur} & & & \\
    \hline
    \hline
    \multirow{3}*{\textit{Arya}} & \textbf{full} & 2,156 & 10.4 & 24 & 5.7 & 137.7 & 80.2 & 3.264 \\
    & \textbf{sty} & 2,180 & 10.4 & 24 & 6.1 & 145.3 & 76.0 & 1.280 \\
    & \textbf{bsl} & 2,180 & 10.4 & 14 & 10.4 & 145.1 & 76.1 & 1.229 \\
    \hline
    \hline
    \multirow{3}*{\textit{Daenerys}} & \textbf{full} & 1,171 & 10.6 & 15 & 6.3 & 93.8 & 139.0 & 1.745 \\
    & \textbf{sty} & 1,185 & 10.6 & 16 & 6.0 & 96.6 & 135.0 & 0.914 \\
    & \textbf{bsl} & 1,185 & 10.6 & 10 & 9.5 & 95.5 & 136.9 & 0.898 \\
    \hline
    \hline
    \multirow{3}*{\textit{Jaime}} & \textbf{full} & 962 & 11.0 & 25 & 6.4 & 153.4 & 71.2 & 1.597 \\
    & \textbf{sty} & 963 & 11.0 & 25 & 6.9 & 172.4 & 65.9 & 0.896 \\
    & \textbf{bsl} & 963 & 11.0 & 15 & 11.1 & 167 & 68.0 & 0.873 \\
    \hline
    \hline
    \multirow{3}*{\textit{Sansa}} & \textbf{full} & 888 & 11.0 & 24 & 5.8 & 139.6 & 91.4 & 1.596 \\
    & \textbf{sty} & 892 & 11.0 & 24 & 6.1 & 146.1 & 87.4 & 0.892 \\
    & \textbf{bsl} & 892 & 11.0 & 15 & 9.7 & 146.2 & 87.4 & 0.881 \\
    \hline
    \hline
    \multirow{3}*{\textit{Theon}} & \textbf{full} & 650 & 10.9 & 16 & 6.0 & 95.7 & 81.6 & 1.431 \\
    & \textbf{sty} & 655 & 10.9 & 15 & 6.4 & 96.0 & 81.3 & 0.819 \\
    & \textbf{bsl} & 655 & 10.9 & 9 & 10.9 & 98.3 & 79.4 & 0.825 \\
    \hline
    \end{tabular}
\end{table}

As can be seen, the number of candidate \textsc{lsu}s differs from one
character to the other: for each character, the only \textsc{lsu}s
considered are those where he/she is verbally active in order to
center the summary on this specific character. Moreover, the
style-based and baseline summaries rely on slightly more candidate
\textsc{lsu}s than the full ones: a few scenes only containing
\textsc{lsu}s with isolated utterances of the character with no
hypothesized interlocutor were discarded when building the full
summaries, but not when constructing the other two types: social
relevance can not apply to these \textsc{lsu}s hypothesized as
soliloquies, but music and shot size may still make them stylistically
salient.

The final summaries turn out to be quite short, ranging from 1:30 to
2:50 minutes, resulting in very high compression rates: the whole
story of a character during 50 one-hour episodes is summarized in
about two minutes. When summarizing the storyline of important
characters, like Daenerys Targaryen, the compression rate may be much
higher than when summarizing the storylines of characters that are not
as important. The total time of the summary is actually dependent on
the number of narrative episodes resulting from the segmentation of
the storyline based on the character's evolving social network:
characters with fast-evolving social environments, going through more
narrative episodes may therefore need longer summaries than possibly
more important characters involved in fewer narrative episodes. Moreover, the
duration of the full summary is sometimes not as long as the
style-based and baseline ones. When building full summaries, the
\textsc{lsu}s are selected separately in each narrative episode, until
the limit of 25 seconds is reached for each one. This may result in a
cumulative loss of a few seconds with respect to the total time
available (25 seconds $\times$ number of narrative episodes). For both
other types of summaries, with \textsc{lsu}s selected from a single,
global set, the loss is usually not as important and the global time
limit is nearly reached.

Not surprisingly, the criterion used when building the full and
style-based summaries by applying Agorithm~\ref{algo:select_lsu}
results in summaries consisting of shorter sequences than the baseline
summary: while the candidate \textsc{lsu}s last a bit more than 10
seconds in average (fourth column in Table~\ref{tab:summ_features}),
the duration of the selected ones (sixth column) in the \textbf{full}
and \textbf{sty} summaries, based on an optimal ratio between
relevance and duration, is very close to the lower bound of 5 seconds
put on the duration of the candidate \textsc{lsu}s (see
Subsection~\ref{subsec:extraction}). On the opposite, the sequences
inserted in the \textbf{bsl} summaries are almost twice as long and
very close to the average duration of the candidate ones.

As can be seen in the last column, the computation time for dynamically generating the summaries on a personal laptop (Intel Xeon-E3-v5 \textsc{cpu}) remain quite low, once \textsc{lsu}s, shot size, background music, along with the dynamic network of interacting characters, have been pre-computed once and for all. Such computation times turn out to be well-suited for the on-demand summary generation scenario we described in Section~\ref{sec:intro}. Nonetheless, though based upon the same algorithm~\ref{algo:select_lsu} (time complexity in $O(n^2)$), full summaries need a bit more time to be generated than the other two types: as explained above, in full summaries, the \textsc{lsu}s are selected separately in each narrative episode, whereas they are globally considered as a single set of possible candidate when building the style-based and baseline summaries.

Finally, the three summaries may overlap. Table~\ref{tab:summ_overlap}
reports for each of the five considered characters the overlapping
time, expressed in \%, between the three summary types.

\begin{table}[h!]
  \caption{\label{tab:summ_overlap} Overlapping time (in \%)
      between the three summaries for each considered character.}
  \vspace{2mm}
  \centering
  \begin{tabular}{|c|r|r|r|r|r|}
    \hline
    \multirow{2}*{\textit{Summaries}} & \multicolumn{5}{c|}{\textit{Character}} \\
    \cline{2-6}
    & Arya & Daenerys & Jaime & Sansa & Theon \\
    \hline
    \textbf{bsl / full} & 4.85 & 3.12 & 1.70 & 2.03 & 4.85 \\
    \textbf{bsl / sty} & 12.62 & 9.80 & 0.75 & 9.11 & 6.28 \\
    \textbf{full / sty} & 30.26 & 32.61 & 32.80 & 35.54 & 36.07 \\
    \hline
  \end{tabular}
\end{table}

    As expected, the overlapping time between the \textbf{full} and
\textbf{sty} summaries on the one hand, and the \textbf{bsl} summary
on the other hand, remain quite low and randomly ranges from 2\% to
12\%, depending on the considered character. In contrast, the overlapping time between the \textbf{full} and \textbf{sty} summaries ranges by construction from 30\% to 36\%: as stated above, \textbf{full} summaries are partially based on \textbf{sty} ones, by additionnaly incorporating socially relevant sequences. The following evaluation protocol is all about measuring, in a contrastive manner, the subjective impact of the consideration of the social content of the plot, in addition to the style of the narrative.

\subsection{Evaluation Protocol}
\label{subsec:protocol}

The users were asked to rank, for each character, the three summaries
according to the two usual criteria used in subjective evaluation of
summaries, \textit{informativeness} and \textit{enjoyability}, but
reformulated as follows according to the specific use case we target:

\textit{1. Which of these three summaries reminds you the most the
  character's story?}

\textit{2. Which of these three summaries makes you the most wanting
  to know what happens next to the character?}

The same questions were asked for their last choice, resulting in a
full ranking of the three summaries for each character. In addition,
the participants were asked to motivate in a few words their
ranking. The participants were allowed not to answer the ranking
questions if too unsure.

No restriction was put on the number of possible viewings of the three
summaries: a passive, first viewing of the summaries was actually
expected to be needed to help the viewer to remember the main steps of
the character's storyline; a second, informed viewing, possibly
cursive, was then expected to be needed to finely compare the
summaries according to the two criteria (respectively referred to as
``\textit{best as recap?}'' and ``\textit{best as trailer?}'' in the
rest of the article). About 25\% or the participants in average
admitted to need several viewings to rank the three summaries
according to the two criteria.

\subsection{Results}
\label{subsec:results}

For each character's storyline, the best summaries according to those
of the participants who had watched all five past seasons of
\textsc{GoT} are reported in Table~\ref{tab:results}, both
as a proportion (denoted ``\%'') and a number (denoted ``\textit{\#}'') of
participants.

\begin{table}[h]
  \renewcommand{\arraystretch}{1.0}
  \caption{\label{tab:results} For each character's storyline, 
    best summary according to the participants who had watched all
    five past seasons of \textsc{GoT}.}
  \centering
  \begin{tabular}{|c|c|rrr|rrr|}
    \hline
    \multirow{2}*{Character} & \multirow{2}*{Votes} & \multicolumn{3}{c|}{\textit{Best as recap?}} & \multicolumn{3}{c|}{\textit{Best as trailer?}} \\
    \cline{3-8}
    & & \textbf{full} & \textbf{sty} & \textbf{bsl} & \textbf{full} & \textbf{sty} & \textbf{bsl} \\
    \hline
    \hline
    \multirow{2}*{\textit{Arya}} & \% & \textbf{70.9} & 9.3 & 19.8 & \textbf{57.1} & 16.7 & 26.2 \\
    & \textit{\#} & \textit{\textbf{61}} & \textit{8} & \textit{17} & \textit{\textbf{48}} & \textit{14} & \textit{22} \\
    \hline
    \hline
    \multirow{2}*{\textit{Daenerys}} & \% & \textbf{35.8} & 32.8 & 31.3 & 18.2 & \textbf{47.0} & 34.8 \\
    & \textit{\#} & \textit{\textbf{24}} & \textit{22} & \textit{21} & \textit{12} & \textit{\textbf{31}} & \textit{23} \\
    \hline
    \hline
    \multirow{2}*{\textit{Jaime}} & \% & \textbf{41.5} & 40.0 & 18.5 & 35.9 & \textbf{43.8} & 20.3 \\
    & \textit{\#} & \textit{\textbf{27}} & \textit{26} & \textit{12} & \textit{23} & \textit{\textbf{28}} & \textit{13} \\
    \hline
    \hline
    \multirow{2}*{\textit{Sansa}} & \% & \textbf{47.7} & 33.8 & 18.5 & \textbf{58.5} & 20.0 & 21.5 \\
    & \textit{\#} & \textit{\textbf{31}} & \textit{22} & \textit{12} & \textit{\textbf{38}} & \textit{13} & \textit{14} \\
    \hline
    \hline
    \multirow{2}*{\textit{Theon}} & \% & 15.6 & \textbf{45.3} & 39.1 & 14.3 & \textbf{57.1} & 28.6 \\
    & \textit{\#} & \textit{10} & \textit{\textbf{29}} & \textit{25} & \textit{9} & \textit{\textbf{36}} & \textit{18} \\
    \hline
    \hline
    \textbf{average} & \% & \textbf{42.3} & 32.2 & 25.4 & 36.8 & \textbf{36.9} & 26.3 \\
    \hline
	\end{tabular}
\end{table}

As can be seen in Table~\ref{tab:results}, whatever the ranking criterion, baseline summaries never obtained majority votes (bold entries in the table): for three of
these summaries (Arya, Jaime, Sansa), the scores remain quite low. In
some cases, participants chose the baseline summary because of the
length of the sequences (about 10 seconds, twice as much as in the
other summary types), perceived as more appropriate to fully
understand and remember the sequences. However, setting the lower
bound of the admissible \textsc{lsu}s to a higher value would first
have resulted in too long summaries: many users reported during
our study that 2-3 minutes was the maximum duration they could
tolerate for video summaries of this type. Furthermore, increasing the sequence duration would have made the user's feedback trickier to interpret: our
method aims at showing both socially relevant and stylistically
salient verbal interactions between characters, but is not sensitive
to their linguistic content. By keeping the sequences short enough, it
is easier, though still challenging, to rely on the user's feedback to
assess our method.

\begin{figure}[h]
    \centering 
  \begin{tabular}{cc}
    \subfloat[\textbf{sty}] {
      \includegraphics[width=.46\textwidth]{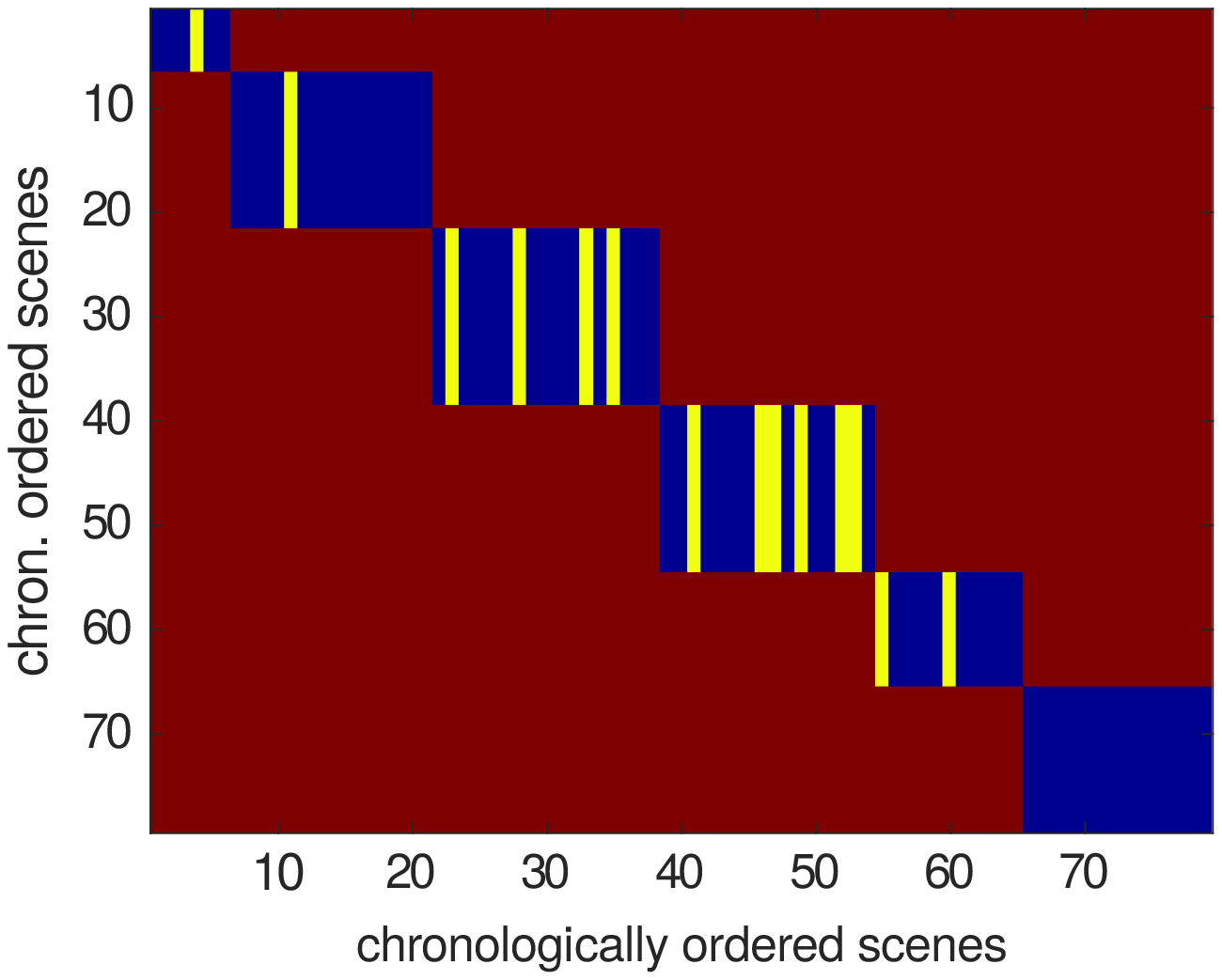}
      \label{subfig:arya_sty_sel}
    }
    &
    \subfloat[\textbf{full}] {
      \includegraphics[width=.46\textwidth]{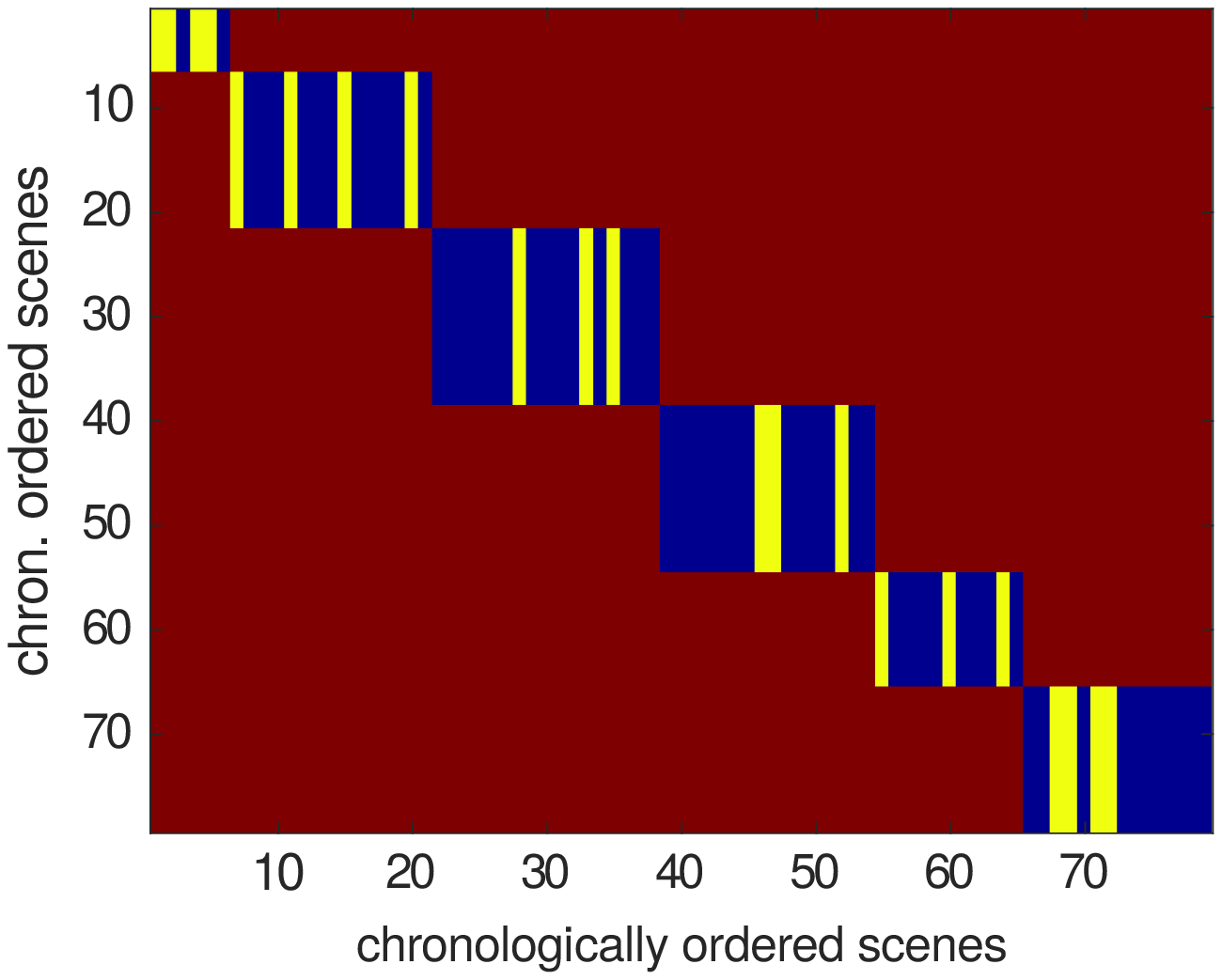}
      \label{subfig:arya_full_sel}
    }
    \end{tabular}
  \caption{\label{fig:arya_sel_dist} Distribution of the selected \textsc{lsu}s (yellow vertical lines) over the narrative episodes (blue boxes) for Arya's storyline summaries: style-based (\textbf{sty}, \ref{subfig:arya_sty_sel}) and full (\textbf{full}, \ref{subfig:arya_full_sel}). Figures available at \href{https://doi.org/10.6084/m9.figshare.7973540}{10.6084/m9.figshare.7973540} (CC-BY license).}
\end{figure}

Nonetheless, concise summaries with short, socially diverse sequences
extracted from every narrative episode were globally well
perceived, and turn out to be worth the slight extra computation time reported in Table~\ref{tab:summ_features}. For 4 out the 5 characters targeted, the full summary was
selected as the most efficient recap, in some cases by far: Arya's
story full summary was considered as the best recap by 70.9\% of the
participants. Interestingly, the two other summaries turn out to miss some major narrative episodes. Fig.~\ref{fig:arya_sel_dist} shows the distribution of the selected \textsc{lsu}s in both style-based (\ref{subfig:arya_sty_sel}) and full (\ref{subfig:arya_full_sel}) summaries of Arya's storyline.

As can be seen, the last narrative episode (last blue box from left to
right on both figures), is not represented in the style-based summary,
in contrast to the full summary. As a result, the last, fifth season
of \textit{Game of Thrones} is not represented in Arya's style-based
summary, which was perceived as ``incomplete'' by many
participants. Moreover, stylistic saliency is likely to be
underrepresented in short narrative episodes, as in the first one (first
blue box from left to right on both figures),
resulting in incomplete summaries unable to capture the whole dynamics
of the character's storyline.

Furthermore, for 4 out of the 5 characters, the full summaries obtain
higher scores when judged as recaps than when judged as trailers. In
some cases, the difference is impressive: whereas 35.8\% of the
participants rank the full summary of Daenerys' story as the best
recap, they are only 18.2\% to rank it as the best trailer. Such a
difference of ranking when switching from the ``recap'' to the
``trailer'' criterion globally benefits the style-based summaries,
more appreciated as trailers than as recaps for 4 characters out of
5. For 3 characters, such style-based summaries even obtain a majority
vote according to the ``trailer'' criterion.

However, some of the votes were sometimes
unexpected. First, the three summaries of Daenerys' storyline obtain
roughly similar scores when evaluated as recaps, without clear
advantage for the full summary: Daenerys is a key-character of
\textsc{GoT}, often named ``Mother of Dragons'' from the fact she owes
three dragons. Many participants, as can be seen from the short
explanations they gave to motivate their ranking, turned out to focus
on this aspect of Daenerys to assess the summaries: absent from the
full summary, Daenerys' dragons are heard in the style-based one, and
by chance seen in the baseline one. Such a criterion was sometimes used to discard
the full summary, though being the only one that captured her crucial
meeting with Tyrion Lannister. The scores obtained by Theon's summary
were also surprising, with quite low scores for the full summary,
probably penalized by a baseline summary rather semantically
consistent and convincing, though somehow incomplete (``Fall of
Winterfell'' and ``Final Reunion with Sansa'' missing).

Nonetheless, on the one hand, the results globally strengthen our
``plot modeling'' approach when it comes to summarize the dynamics of
a character's storyline over dozens of episodes. On the other hand,
the stylistic features we used to isolate salient sequences remain too
hazardous when used on their own for capturing the whole character's
storyline, but turn out to be valuable to make viewers feel like
viewing what comes next.

\section{Conclusion and Perspectives}
\label{sec:conclu}

In this paper, we described and evaluated a way to automatically
generate character-oriented summaries of \textsc{tv} serials from partially annotated data. We first
described a method for extracting consistent candidate sequences for
later, possible insertion into the final character's storyline
summary, before detailing a weighting scheme of their relevance. On
the one hand, the relevance of each candidate sequence, based on a
step of pre-segmentation of the character's storyline into narrative
episodes, is expressed in terms of social relevance. The summary is
then designed so as to focus on the most typical relationships of the
character at each step of his/her storyline. On the other hand, the relevance of each candidate sequence can
also be expressed in terms of
stylistic saliency. We specifically focused on shot size and
background music, as mid-level features commonly used by filmmakers to
emphasize the importance of some specific sequences. We evaluated this
summarization framework by performing a large scale user study in a
real case scenario and obtained promising results. The social network
perspective that we introduced results in content-covering summaries that
the viewers perceived as effective recaps of the complex \textsc{tv}
serial plots; in addition, the use of background music and shot size
for supporting the engaging, revival effect expected from such
summaries turned out to be relevant to the users we polled.

In future work, we would like to generalize our feature choices, both
content and style-related, to other \textsc{tv} serial genres, and
further investigate the empathetic relationship that our
character-oriented summaries could revive between the viewer and
his/her favorite character(s): the cold-start phenomenon that the
season trendlines of IMDb ratings exhibit on
Fig.~\ref{fig:GoT_ratings} not only depends on the viewer's cognitive
disengagement, but probably also on an emotional disaffection that
character-oriented summaries can handle properly.

\begin{acknowledgements}
  This work was supported by the Research Federation \textit{Agorantic}, Avignon University.
\end{acknowledgements}

\newpage



%
%

\end{document}